\documentstyle[pstricks,aps,floats,psfig,pst-plot]{revtex}

\begin{document}

\twocolumn[\hsize\textwidth\columnwidth\hsize\csname
@twocolumnfalse\endcsname

\title{\vspace*{-1cm}\hfill
%{\tt Submitted to Phys.~Rev.~Letters}
       \vspace{0.5cm}\\
Formation and stability of self-assembled coherent islands  
in highly mismatched heteroepitaxy}

\author{L. G. Wang, P. Kratzer, M. Scheffler, and N. Moll\cite{author}}
\address{
Fritz-Haber-Institut der Max-Planck-Gesellschaft, Faradayweg 4-6,
D-14195 Berlin-Dahlem, Germany}

\date{\today}
\maketitle
\begin{abstract}

We study the energetics of island formation in Stranski-Krastanow 
growth within a parameter-free approach. It is shown that an optimum 
island size exists for a given coverage and island density if changes 
in the wetting layer morphology after the 3D transition are properly 
taken into account. Our approach reproduces well the experimental 
island size dependence on coverage, and indicates that the critical 
layer thickness depends on growth conditions. The present study provides 
a new explanation for the (frequently found) rather narrow size 
distribution of self-assembled coherent islands.

\vspace{0.2cm}

\end{abstract}
\vskip2pc] 

The surface morphology of overlayers in  
heteroepitaxial growth
has attracted intense interest because of its  
importance for basic science and applications in  optoelectronic devices. 
Experiments \cite{Mo90,Leonard94,Moison94,Heitz97,Carlsson94}  
showed that heteroepitaxy in systems with a  
lattice constant difference $\geq$ 2\%,
such as InAs/GaAs\cite{Leonard94,Moison94,Heitz97}, Ge/Si\cite{Mo90}
and InP/InGaP\cite{Carlsson94}, 
 follows the so-called 
Stranski-Krastanow growth mode\cite{Stranski39}: 
three-dimensional (3D) dislocation-free
(so-called coherent) islands form on top of the wetting layer.
These small coherent islands are often found to have a very narrow size 
distribution\cite{Leonard94,Moison94,Heitz97}
and are promising to be used in  
quantum dot light emitting diodes (LEDs) and lasers.

It is commonly agreed that the energetics of strain relief plays a key role 
in the growth process:
Islands form, instead of a uniformly strained, epitaxial film, 
because the gain of elastic relaxation energy in an island 
overcompensates the cost due to the increased surface energy 
by islanding. 
It is tempting to attribute the observed island size distribution to a 
minimum of the free energy of the system. 
However, an equilibrium theory with only two energetic contributions, a 
positive one from the island surface energy 
($E \sim V^{2/3}$, $V$ is the quantum dot volume) 
and a negative one from the elastic relaxation energy 
($E \sim V$), fails to predict a 
finite equilibrium size of the islands.
Instead, the energy gain from strain relief always prevails for sufficiently 
high coverages, rendering larger islands more stable than smaller ones.  
In order to cope with this difficulty, 
several additional effects, e.g. contributions from intrinsic surface stress 
or from interactions between islands\cite{Shchukin95,Daruka97}, 
have been invoked. 
Priester and Lannoo\cite{Priester95} proposed a 
mechanism in which 2D platelets act as 
precursors for the formation of 3D coherent islands, thus determining 
their size.  
Most recently, the observation of island ripening\cite{Lee98} has made it 
doubtful if the islands can be interpreted at all as structures in total 
equilibrium. 

\begin{figure}[b!]
\pspicture(1.8,2.2)
\rput[c](4.3,1.4){
\psfig{figure=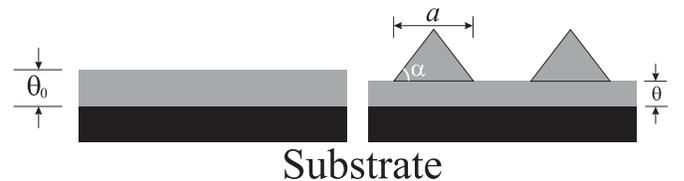,height=2.3cm}}
\endpspicture
\noindent\caption{Schematic illustration of the formation of coherent
islands on the substrate surface.
$\theta_0$ and  $\theta$ are the nominal coverage and
wetting layer thickness, respectively. $\alpha$ is the tilt angle
of island facets and $a$ is the island base length.}
\label{dot}
\end{figure}

In this Letter, we show that the narrow size distribution of the 
coherent islands 
can be understood 
as the result of the system being  trapped in a 
{\em constrained} equilibrium state, where the size is 
determined by the island density and the nominal coverage.
In the constrained equilibrium theory, the existing nuclei grow to a 
size determined by  
the energetic balance that governs material transport between 
the wetting layer and the islands. 
This allows us to derive an 
optimum islands size for a fixed coverage and island density 
from a parameter-free approach.
The elastic energy in both the islands and the substrate 
is calculated within continuum elasticity theory. 
Using density-functional theory within the 
local-density approximation, 
accurate surface energies  are obtained with the help of 
the plane-wave pseudopotential method\cite{Bock97} for both 
the island facets and the wetting layer.  
Previous studies\cite{Chen96,Tersoff93} had missed the latter contribution, 
assuming that the surface energy of the wetting layer 
would be unchanged by the 3D transition.
Our theory reproduces very well experimental data for 
the island size  dependence on coverage.
In particular, we improve over previous approaches by 
showing that the selectivity of growth of a certain island size 
can be explained without invoking delicate elastic edge effects or 
island interactions\cite{Shchukin95,Daruka97}. 
Furthermore, we demonstrate how the critical layer thickness depends on  
growth conditions,  settling this long-standing issue. 

\begin{table}
\caption{ Surface energies $\gamma_{\rm f}$ and surface stresses 
$\sigma_x$, $\sigma_y$ for InAs surface reconstructions 
with the chemical potential $\mu_{\rm {As}}=\mu_{\rm {As(bulk)}}-0.2$eV.}
\begin{tabular}{d c c c c}
surface &  & $\gamma_{\rm f}$ & $\sigma_x$ & $\sigma_y$  \\
       &   &  (meV/\AA$^2$) & (meV/\AA$^2$) & (meV/\AA$^2$) \\ \hline
(110) & cleavage & 41 & 26 & 54 \\
(111) & In vacancy & 42 & 48 & 48 \\
($\bar{1}\bar{1}\bar{1}$) & As trimer & 49 & 92 & 92 \\
\end{tabular}
\label{table1}
\end{table}

We propose a view of the growth process divided in three phases: 
an early nucleation phase which mainly determines the island 
density $n$, a second phase where the islands grow mostly on expense 
of the wetting layer, and a third phase characterized by 
Ostwald ripening.
Since we are mostly interested in island sizes, 
we concentrate on the second phase, and briefly discuss the third phase later.
As long as the wetting layer acts as a source for material, 
%transport towards the islands mediated by mobile surface species, their 
%supersaturation is high, and 
existing nuclei will grow rapidly.  
Hereby the island density $n$ remains constant\cite{Moison94}. We treat it 
as an input to our model noting that the island density is determined by
the growth kinetics.
Furthermore, we assume the islands have identical shape and volume $V$. 
%make use of the fact that nucleation leads to a 
%fairly regular spatial distribution of nuclei with roughly similar size, 
%which allows us to approximate the islands by identical objects of 
%volume $V$.
In the following, we discuss the island size in terms of a constrained 
thermodynamic equilibrium between islands and wetting layer, for a 
fixed island density. 
Although our approach is not limited to a certain system, here we 
consider, as an example, the strained \{110\} pyramidal shaped, 
dislocation-free InAs islands  with a square base (area $a^2$) 
on the GaAs(100) surface (with a wetting layer). 
We will also discuss the $\{111\}/\{\bar{1}\bar{1}\bar{1}\}$ faceted 
pyramidal islands  later. 
We choose this 
system with a lattice mismatch $\Delta a \approx$ 7\%, 
because a large number of experimental and theoretical studies 
have been done\cite{Leonard94,Moison94,Heitz97,Priester95}. 
The real island shape may be more complex, 
but the simple island shape used here should still allow us   
to capture the important features of the island formation 
(see below and Ref.\cite{Tersoff93}).
%\cite{Tersoff93}).
% such as size and wetting layer thickness.  
%We assume the islands are regularly located on the substrate 
%surface.
Figure 1 schematically illustrates the 
island formation on the substrate surface.  
$\theta_0$ and  $\theta$ are the nominal coverage and 
wetting layer thickness after island formation, respectively.    
We omit the interaction between islands 
(as we will discuss below this is a 
very good approximation in our case), 
 as well the energies of edges and corners. 
%we say this once more, later, thus not needed here
%
%As will be seen in Fig.3, the energy contribution by edges 
%is very small compared to other contributions. 
The entropic contribution
to free energy is also neglected. 

The total energy gain per unit volume of a single 
island can be expressed as

%$$ nE_{tot}=nV(\epsilon^{el}_{is}-\epsilon^{el}_{film})
%+n(S\gamma_{f}-a^2\gamma_{wl})
%+(1-na^2)*[\gamma_{wl}^\prime(\theta)-\gamma_{wl}]      $$

\begin{eqnarray}
\nonumber  E_{\rm tot}/V=\epsilon^{\rm el}_{\rm is}
-\epsilon^{\rm el}_{\rm film} 
+[S\gamma_{f}-a^2\gamma_{\rm wl}(\theta_0)]/V  \\  
+(1/n-a^2) \times [\gamma_{\rm wl}(\theta)-\gamma_{\rm wl}(\theta_0)]/V  
\end{eqnarray}

\noindent where $\epsilon^{\rm el}_{\rm is}$ and 
$\epsilon^{\rm el}_{\rm film}$ are 
the elastic energy densities of 
island and uniformly strained film. 
The third term describes the change in surface energy 
due to the island,  
with $\gamma_{\rm f}$ being the surface energy of the island 
facets and $S$ their area. 
The fourth term accounts for the thinning of the part 
of the wetting layer which 
feeds the island.  $\gamma_{\rm wl}(\theta_0)$, 
$\gamma_{\rm wl}(\theta)$ are the formation energy
%(including the interface energies)
of the wetting layer as a function of 
%the nominal coverage $\theta_0$ and 
its thickness $\theta$, measured relative to InAs bulk kept at the GaAs lattice
constant. This allows us to introduce the elastic contribution 
to the formation energy as a uniquely defined separate term  
 $\epsilon^{\rm el}_{\rm film}$, similar to earlier work\cite{Moll98}.
% respectively.   
% of wetting layer that is equilibrium with the islands 
%and of nominal coverage layer. In the case that nominal
% coverage is larger 
%than the critical coverage, then $\gamma_{wl0}$=$\gamma_{wl}$.  
From mass conservation, the volume of an island, $V$, is given by 
$V={1\over 6}a^3 \tan \alpha={1\over n}(\theta_0-\theta)L$, 
where $\alpha$ and $L$ are the 
tilt angle of island facets and the monolayer (ML) thickness, respectively.
Equation (1) holds true as long as the island contains 
a sufficiently high number of atoms 
({\em e.g.} \hbox{5 000} atoms) 
because elasticity theory is applicable and the reconstructions 
 on the facets are completed\cite{Moll98}.

\begin{figure}[t!]
\pspicture(4.5,3.9)
\rput[c](4.5,2.4){
\psfig{figure=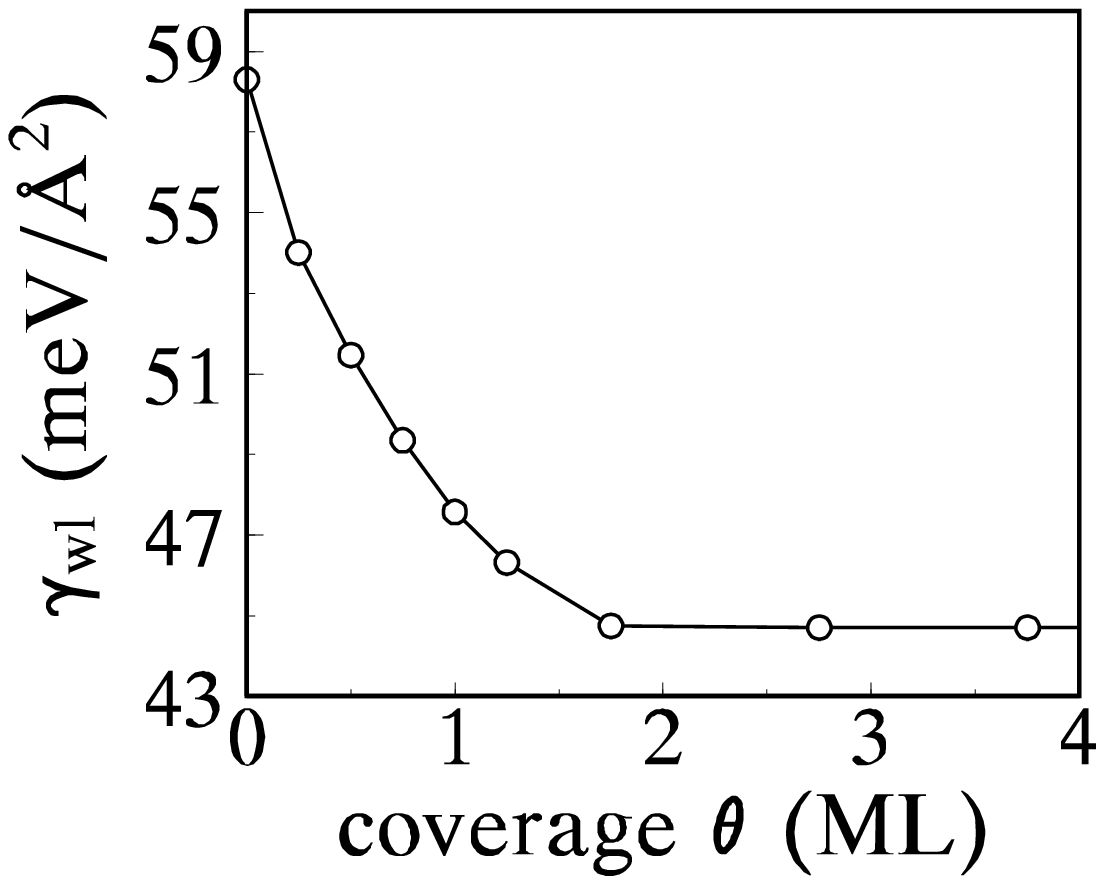,height=7.0cm}}
\endpspicture
\noindent\caption {Formation energy of the wetting layer as a function of 
thickness $\theta$, defined by $\gamma_{\rm wl}A = E^{\rm tot}
 - \mu_{\rm GaAs(bulk)}N_{\rm Ga} $ 
$- \mu^{\rm strained}_{\rm InAs(bulk)}N_{\rm In}
 - \mu_{\rm As}(N_{\rm As} - N_{\rm Ga} - N_{\rm In})$, where $A$ is 
the surface
area and $N_i (i=$As, Ga, In) are the number of particles of 
the species $i$  in the supercell. E$^{\rm tot}$ is the total 
energy of the supercell. 
$\mu_{\rm GaAs(bulk)}$ and $\mu^{\rm strained}_{\rm InAs(bulk)}$ 
are the chemical potential of GaAs bulk and of strained InAs bulk with 
the theoretical equilibrium lattice constant of GaAs bulk. 
From various configurations with $N_{\rm In}=8\/ \theta$ In atoms 
per $(2 \times 4)$ surface unit cell, those with the lowest formation 
energy are presented.
}
\label{surf}
\end{figure}

In order to obtain accurate values for 
the surface energies and intrinsic surface stresses, 
these are computed using slab models of the surfaces with the help of the 
pseudopotential plane-wave method\cite{Bock97}.
After optimizing the atomic geometries 
using consistently calculated forces on the atoms, 
the total energies of the slabs are computed, and 
the formation energies of various surfaces are obtained by 
subtracting the calculated total energy of an appropriate 
amount of bulk material. 
We further take into account  
the surface stress contribution to the surface energy up to the
linear term \cite{Moll98}, and a term proportional to the chemical potential 
$\mu_{\rm As}$ of the environment in case of non-stoichiometric surfaces.
Since epitaxial growth is mostly performed 
under As-rich conditions, 
all surface energies are evaluated close to equilibrium 
with bulk arsenic ({\em i.e.} $\mu_{\rm As}=\mu_{\rm As(bulk)}-0.2$eV). 
For each facet, we have selected 
the reconstruction with the lowest energy  
from several candidates\cite{Moll98}.
For the wetting layer, we consider the $\beta2(2\times4)$ 
reconstruction which is usually found on GaAs(001) and InAs(001) 
surfaces under moderately As-rich conditions\cite{Yamagu95}.
The results are given in Table I and Fig. 2. 
%We can see that the surface energy of the wetting layer 
%decreases with increasing wetting layer thickness;  it 
%converges to the value of pure InAs(001) $\beta2(2\times4)$ 
%at about 1.75 ML.

\begin{figure}[t!]
\pspicture(4.5,5.5)
\rput[c](4.5,3.0){
\psfig{figure=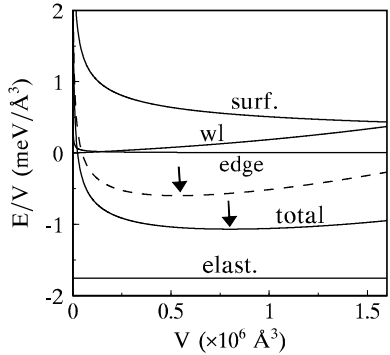,height=8.3cm}}
\endpspicture
\noindent\caption {Total energy gain by islanding and various energy
contributions (solid lines) for  $n=10^{10} {\rm cm}^{-2}, \theta_0=1.8$ ML.
The dashed line is the total energy gain for  $n=10^{10} {\rm cm}^{-2},
\theta_0=1.5$ ML. The arrows mark the minima of the curves.}
\label{etot}
\end{figure}

The elastic energy is calculated within continuum elasticity theory 
using the experimental elastic moduli to describe the elastic properties
of both the island and the substrate. 
For the island plus a 240 {\AA} thick slab representing
the substrate, a finite-element approach is applied to 
solve the nonlinear elasticity problem\cite{Moll98,Pehlke97}.
We calculate the elastic energy for 
several island shapes (different 
tilt angles $\alpha$ of island facets) 
with a fixed island volume.
For each particular shape, we can extract $\epsilon^{\rm el}_{\rm is}$
from the observed linear scaling relation with the island volume.
The elastic energy density $\epsilon^{\rm el}_{\rm film}$
of the uniformly strained film can be obtained by extrapolating 
the results for the islands to $\alpha$=0, and we find 
it is in very good agreement with the value from linear elasticity 
theory.
%Both the island and a sufficiently thick slab representing 
%the substrate 
%are divided into small irregularly shaped tetrahedra. 
%The displacement field is tabulated on the vertices 
%of this partitioning.
%Within each tetrahedron the linear interpolation of 
%the displacement field is uniquely determined by the values at the four 
%corners of the tetrahedron. The total elastic energy 
%is calculated by summing the elastic energy density within each 
%tetrahedron, which is a function of local strain, times the volume 
%of the tetrahedron over all tetrahedra. This expression 
%is iteratively minimized with respect to the displacement field. 
%This procedure is repeated for finer and finer partitioning 
%of the volume and the results are finally extrapolated to fineness 
%equal to zero.
%Details of the calculations have been presented 
%elsewhere\cite{Moll98,Pehlke97}.  
%The calculated elastic energy depends on 
%the island shape (the tilt angle $\alpha$ of island facets). 
%The calculations also show that 
%the top of the pyramidal island is almost fully relaxed\cite{Moll98,Pehlke97}.
% and the 
%elastic energy is almost completely stored in the base and 
%substrate\cite{Moll98}.

In Fig.~3, we show the various energy contributions 
and the total energy gain per volume 
for $n=10^{10}$ cm$^{-2}, \theta_0=1.8$ ML.
The elastic relaxation energy (the first and second term in Eq.(1)) 
is negative due to strain relief and 
scales linearly with the island volume. Surface energy 
(the third term in Eq.(1)) is a cost, and therefore 
its contribution is positive. The wetting layer energy contribution 
(the fourth term in Eq.(1))  
is also positive and depends complexly on the island volume, 
island density and coverage.
We also show the energy contribution of the edges in Fig.~3, which 
becomes negligible compared to the   
other contributions for a large island (estimated as in Ref.\cite{Moll98}).
%\cite{Moll98}).
It is important that an energy minimum exists in the 
total energy gain curve. This indicates an optimum island size 
can be obtained under certain growth conditions.  
% (coverage, substrate temperature and flux rate).
The minimum in Fig.~3 corresponds to an island with about \hbox{38 000} atoms, which  compares 
reasonably well with typical experimental values (between \hbox{20 000}\cite{Moison94} and 
\hbox{50 000} atoms per island\cite{Leonard93}).
The quite uniform islands prior to ripening  observed in Ref.\cite{Lee98} lend  further support to the 
existence of an optimum island size.   
%
% It cannot be attributed to island ripening, because ripening broadens the 
% distribution, as you say later.
% Also, the explanation of Priester and Lanoo is only valid when 
% complete thermodynamic equilibrium is considered, which is not the 
% case here.
%
%The narrow island size distribution\cite{Leonard94,Moison94,Heitz97}  
%might be attributed to  
%the island ripening (see later discussions and Ref.[10])  
%or/and the finite temperature effect\cite{Priester95}. 
 However, the  
island size strongly depends on 
the island density.
Fig.4(a) shows that the equilibrium island volume $V$ 
%(or, likewise, the island half-base $R \approx 8.68n^{-0.27}$ nm)
is a hyperbolic 
function of island density $n$. 
%For $n=2.5\times10^{10}$ cm$^{-2}, 
%\theta_0=1.8$ ML, the equilibrium island size
%contains about 20 000 atoms ($R \approx$ 6.78 nm).
%Fig.4(a) also shows that when the island density is quite high 
%varying the island density does not result in a large change of 
%island size.  
As the experimental observations\cite{Solomon95} have shown,
the island radius varies exponentially  with the growth 
temperature.
%, for example, the 
%island radius is doubled when the growth temperature varies from 
%500 $^\circ$C to 540 $^\circ$C. 
This was attributed to a kinetically limited process\cite{Solomon95}. 
It is consistent with the theory presented here, 
since the island density is known to depend strongly 
on the growth temperature\cite{Amar95,Dobbs97}, and thus kinetics 
controls  the growth through controlling the island nucleation density.
Fig.4(b) shows that our theory can reproduce very well the increase
of the island radius with the amount of deposited material
observed in various experiments, by using suitable island densities as input.
The good agreement between theory and
experiment also justifies the neglect of repulsive interactions
between islands in the present study.
A careful check indicates that the distances between the islands
are quite large (larger than 65 nm and 220 nm for the high
[2.3$\times10^{10}$ cm$^{-2}$] and low
[1.9$\times10^{9}$ cm$^{-2}$] island densities, respectively).
The island density $2.3\times10^{10}$ cm$^{-2}$ used to
fit the experimental
results\cite{Polimeni96} agrees well with the experimentally
estimated island density $1.5-2\times10^{10}$ cm$^{-2}$.
%We are also able to reproduce very well the experimental island size 
%dependence on the island density (see Fig.2 in Ref.\cite{Ebiko98}) 
%by applying the relation between the island density and the coverage 
%obtained by Leonard {\em et al.}\cite{Leonard94} to our theory.  
%
% I think we should not say that, because we assume that 
% the island density is constant and doesn't change with time. 
%
%It is worthy to notice that if the island density increases during the  
%deposition process (specially for the initial stage of the 
%island formation), one could find the 
%island size does not evidently increase\cite{Leonard94}.   

\begin{figure}[t!]
\pspicture(4.5,7.1)
\rput[c](4.5,4.6){
\psfig{figure=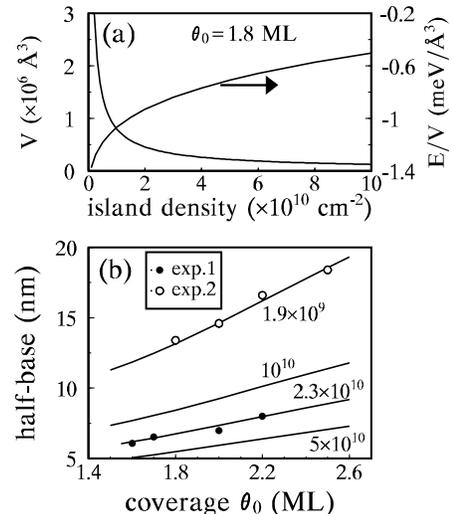,height=10.0cm}}
\endpspicture
\noindent\caption {The dependence of the optimum island size
(volume and half-base)
on the island density and the nominal coverage.
In (a), the total energy gains for the optimum island size
in various island densities are also shown.
The experimental values in (b) are taken from Ref.[21]
($\bullet$) and estimated from Ref.[3] ($\circ$).}
\label{veq}
\end{figure}

The critical layer thickness is a very interesting issue 
and the reported 
values vary from 1.2 to 2 ML\cite{Polimeni96,Berti97,Bottom98}. 
Our present theory puts us in position to 
discuss the critical layer thickness,  
because the total energy gain depends on the coverage (see Fig.~3). 
When we deposit less material, keeping $n$ fixed,  
the energy minimum rises above zero, {\em i.e.} island formation is 
no longer favorable. 
We take the critical layer thickness as the coverage at which 
the minimum energy equals zero 
(the error $\pm$ 0.01 ML). Our results, in Fig.5, indicate that 
the critical layer thickness varies from 1.20 to
1.79 ML when the island density varies 
from $10^9$ cm$^{-2}$ to 3.5$\times10^{11}$ cm$^{-2}$.
Our theoretical prediction matches the experimentally   
observed range (1.2-2 ML).
We note that an exact determination is difficult, 
because in the experiment various other 
factors may influence the critical layer thickness,  
like more complex island shapes,
details of the growth method and growth conditions 
(e.g. III/V ratio)\cite{Snyder93,Xue97}, 
or a possible correlation between island density and 
coverage\cite{Leonard94}. 

\begin{figure}[t!]
\pspicture(4.5,5.0)
\rput[c](4.5,3.0){
\psfig{figure=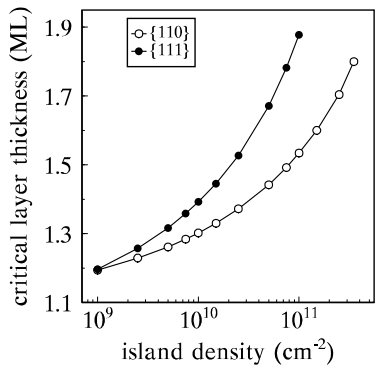,height=7.0cm}}
\endpspicture
\noindent\caption {Critical layer thickness as a function of
the island densities.}
\label{crilay}
\end{figure}

We have also performed a corresponding analysis for the strained 
InAs$\{111\}/\{\bar{1}\bar{1}\bar{1}\}$ pyramidal 
islands. The results show similar behavior as for the 
InAs\{110\} islands  and indicate  that the basic  
features of our model do not depend on shape assumptions. 
The calculations also indicate 
that a larger island tends to be a steeper one, {\em i.e.} 
a $\{111\}/\{\bar{1}\bar{1}\bar{1}\}$ 
faceted pyramidal island,
%whereas for smaller ones the \{110\} facet shape is 
%favored \cite{Moll98,Pehlke97}.  
%We attribute this result 
due to enhanced elastic energy relief
%: a steeper island allows for a larger fraction of its volume to relax
\cite{Moll98,Shchukin96}.
Our study with 
strained $\{111\}/\{\bar{1}\bar{1}\bar{1}\}$ pyramidal islands also shows 
that the critical 
layer thickness somewhat 
depends on the island shape (see Fig.~5).
However, we can still predict trends,
%that the critical layer thickness depends on growth conditions, 
{\rm e.g.} for
high growth temperatures (having a small island nucleation density)
the critical layer is thinner. 
Our study also indicates that in case of a thinner 
critical layer, 
the island embryo should be larger than that for a thicker one. 
This can be understood in terms of a larger energy barrier 
which must be  
overcome by the embryo when nucleating on a thinner 
wetting layer.      

Finally, we briefly comment on the ripening of the islands. 
When no more material is supplied by the wetting layer, 
%the supersaturation of mobile surface species drops quickly.  
%In this regime, 
the island density is no longer constant, because 
smaller islands will dissolve again. 
Allowing the island density to vary, 
we find that larger islands at a lower density are energetically 
preferred (see  Fig.4(a)).
Thus, our theory is in accord with the observed Ostwald ripening\cite{Lee98}. 
However, since noticeable changes in the island size and density 
resulting from
ripening typically take many days, ripening is not important for device 
applications, where the islands are covered by a capping layer after 
a very short growth interruption, and was not observed in 
previous experiments performed on a shorter time scale.
% leave out: no space left
%
%The ripening brings two changes: the island density decreases
%and island size distribution broadens\cite{Lee98}. 
%So the experimental results depend on when the measurements are 
%performed. Our study also indicates the ripening is a intrinsic 
%process ignoring whether or not the dislocations exist in the islands
%since we do not consider the effect of dislocations in our theory. 

In conclusion, we presented  a novel explanation 
for the selection of 
particular sizes of 
self-assembled coherent islands in highly mismatched heteroepitaxy. 
%by taking the wetting layer change into account.
It is possible to select the island size  
by changing the growth conditions and 
the nominal coverage. 
%We predict the well-defined size islands are unstable 
%and will go ripening.
Our theory reproduces very well the 
experimental trends observed in the island growth. 
We have also shown  how the critical layer thickness 
depends on growth conditions and settled this long-standing issue.   

The authors thank E. Pehlke for fruitful discussions. 
%One of authors (L. G. Wang)
%thanks T. Zywietz and J. Neugebauer for the help 
%in using the FHIMD96 code.
This work was supported by Sfb 296 of the  Deutsche
Forschungsgemeinschaft.

\end{document}